\documentclass[]{spie}  
\usepackage[]{graphicx}
\usepackage{acronym}
\usepackage{longtable}
\usepackage{textcomp}
\newcommand{\pb}{\textsc{Polarbear}}
\newcommand{\Pb}{\textsc{Polarbear}}

\newcommand{\etal}{\textit{et al.}}

\newcommand{\figref}{Figure~}
\newcommand{\secref}{Section~}
\newcommand{\tabref}{Table~}

\acrodef{tes}[TES]{Transition Edge Sensor}
\acrodef{snr}[SNR]{Signal-to-Noise Ratio}
\acrodef{hwp}[HWP]{Half-Wave Plate}
\acrodef{sq}[SQUID]{Superconducting Quantum Interference Device}
\acrodef{cmb}[CMB]{Cosmic Microwave Background}
\acrodef{fts}[FTS]{Fourier Transform Spectrometer}
\acrodef{pei}[PEI]{polyetherimid}
\acrodef{uhmwpe}[UHMWPE]{ultra-high molecular weight polyethylene}
\acrodef{ar}[AR]{Anti-Reflection}
\acrodef{lsn}[LSN]{Low-stress silicon nitride}

\title{The bolometric focal plane array of the {\sc \LARGE Polarbear} CMB experiment} 

\author{K. Arnold\supit{a}, 
P.A.R. Ade\supit{b}, 
A.E. Anthony\supit{c},
D. Barron\supit{d} 
D. Boettger\supit{d}, 
J. Borrill\supit{e,f}, 
S. Chapman\supit{g} 
Y. Chinone\supit{k},
M.A. Dobbs\supit{h}, 
J. Errard\supit{ j}, 
G. Fabbian\supit{ j}, 
D. Flanigan\supit{a},
G. Fuller\supit{d}, 
A. Ghribi\supit{a},
W. Grainger\supit{n}, 
N. Halverson\supit{c}, 
M. Hasegawa\supit{k},
K. Hattori\supit{k}
M. Hazumi\supit{k}, 
W.L. Holzapfel\supit{a}, 
J. Howard\supit{a}, 
P. Hyland\supit{i}, 
A. Jaffe\supit{m}, 
B. Keating\supit{d}, 
Z. Kermish\supit{a}, 
T. Kisner\supit{e}, 
M. Le Jeune\supit{ j}, 
A.T. Lee\supit{a,m}, 
E. Linder\supit{n}, 
M. Lungu\supit{a},
F. Matsuda\supit{d},
T. Matsumura\supit{k},
N.J. Miller\supit{d} 
X. Meng\supit{a}, 
H. Morii\supit{k}, 
S. Moyerman\supit{d},
M.J. Myers\supit{a},
H. Nishino\supit{a},
H. Paar\supit{d}, 
E. Quealy\supit{a}, 
C. Reichardt\supit{a},
P.L. Richards\supit{a},
C. Ross\supit{g}, 
A. Shimizu\supit{k} 
C. Shimmin\supit{a}, 
M. Shimon\supit{d}, 
M. Sholl\supit{m}, 
P. Siritanasak\supit{d},
H. Spieler\supit{m}, 
N. Stebor\supit{d}, 
B. Steinbach\supit{a}, 
R. Stompor\supit{j}, 
A. Suzuki\supit{a}, 
T. Tomaru\supit{k}, 
C. Tucker\supit{b}, 
O. Zahn\supit{a,m}
\skiplinehalf
\supit{a}Department of Physics, University of California, Berkeley CA 94720 \\
\supit{b}School of Physics and Astronomy, University of Cardiff \\
\supit{c}Department of Astrophysical and Planetary Sciences, University of Colorado \\
\supit{d}Center for Astrophysics and Space Sciences, University of California, San Diego \\
\supit{e}Computational Cosmology Center, Lawrence Berkeley National Laboratory \\
\supit{f}Space Sciences Laboratory, University of California, Berkeley \\
\supit{g}Dalhousie University \\
\supit{h}Physics Department, McGill University \\
\supit{i}Physics Department, Austin College \\
\supit{j}Laboratoire Astroparticule et Cosmologie (APC), Universite Paris 7 \\
\supit{k}High Energy Accelerator Research Organization (KEK), Tsukuba, Ibaraki, Japan \\
\supit{l}Department of Physics, Imperial College \\
\supit{m}Physics Division, Lawrence Berkeley National Laboratory, Berkeley, CA 94720 \\
\supit{n}Rutherford Appleton Laboratory, STFC
}

\authorinfo{Contact Author: Kam Arnold (karnold@berkeley.edu)}

\begin{document} 
\maketitle 

\begin{abstract}
The \pb\ \ac{cmb} polarization experiment is currently observing from the Atacama Desert in Northern Chile. It will characterize the expected B-mode polarization due to gravitational lensing of the \ac{cmb}, and search for the possible B-mode signature of inflationary gravitational waves. Its 250 mK focal plane detector array consists of 1,274 polarization-sensitive antenna-coupled bolometers, each with an associated lithographed band-defining filter. Each detector's planar antenna structure is coupled to the telescope's optical system through a contacting dielectric lenslet, an architecture unique in current \ac{cmb} experiments. We present the initial characterization of this focal plane.
\end{abstract}



\acresetall
\section{INTRODUCTION}
\label{sec:intro}

\Pb\ is designed to measure the B-mode \ac{cmb} polarization signature due to gravitational lensing and search for the possible B-mode signature of inflation. Kermish \etal\cite{Kermish_SPIE2012}, also at this conference [8452-47], presents the scientific goals and instrumental overview of \Pb. In this proceeding, we present the details of the antenna-coupled bolometric focal plane detector array used in the \pb\ experiment. 

To achieve the high sensitivity and stringent control of systematic instrumental effects required by \pb's scientific goals, there are several important detector design criteria: The individual detectors must couple optical power from the free-space telescope optics to the bolometer with minimal loss. The spectral band of the detectors, centered at 148 GHz, must be designed to utilize the atmospheric window around that frequency. The detectors must be dual-polarization, with diffraction-limited spatial response that are symmetric between polarizations. The bolometer time constants must be such that they can measure the \ac{cmb} structure in the sky as the telescope scans across it. And the bolometer sensitivity must be background-limited, recognising that the statistical variations in the photon flux from the atmosphere are the fundametal limit to detector sensitivity for ground-based observation of the \ac{cmb}.

Section \ref{sec:fparc} of this proceeding goes over the design of the focal plane array of detectors and how it addresses these goals, while Section \ref{sec:fab} gives the details of the detector fabrication in the Berkeley Nanolab. Section \ref{sec:valid} presents the results of tests conducted to validate the focal plane performance. Section \ref{sec:future} concludes with some comments about the future of this technology and of the \pb\ experiment.

\section{Focal plane array architecture} 
\label{sec:fparc}

   \begin{figure}
   \begin{center}
	\begin{tabular}{cc}
   \includegraphics[width=2.5in]{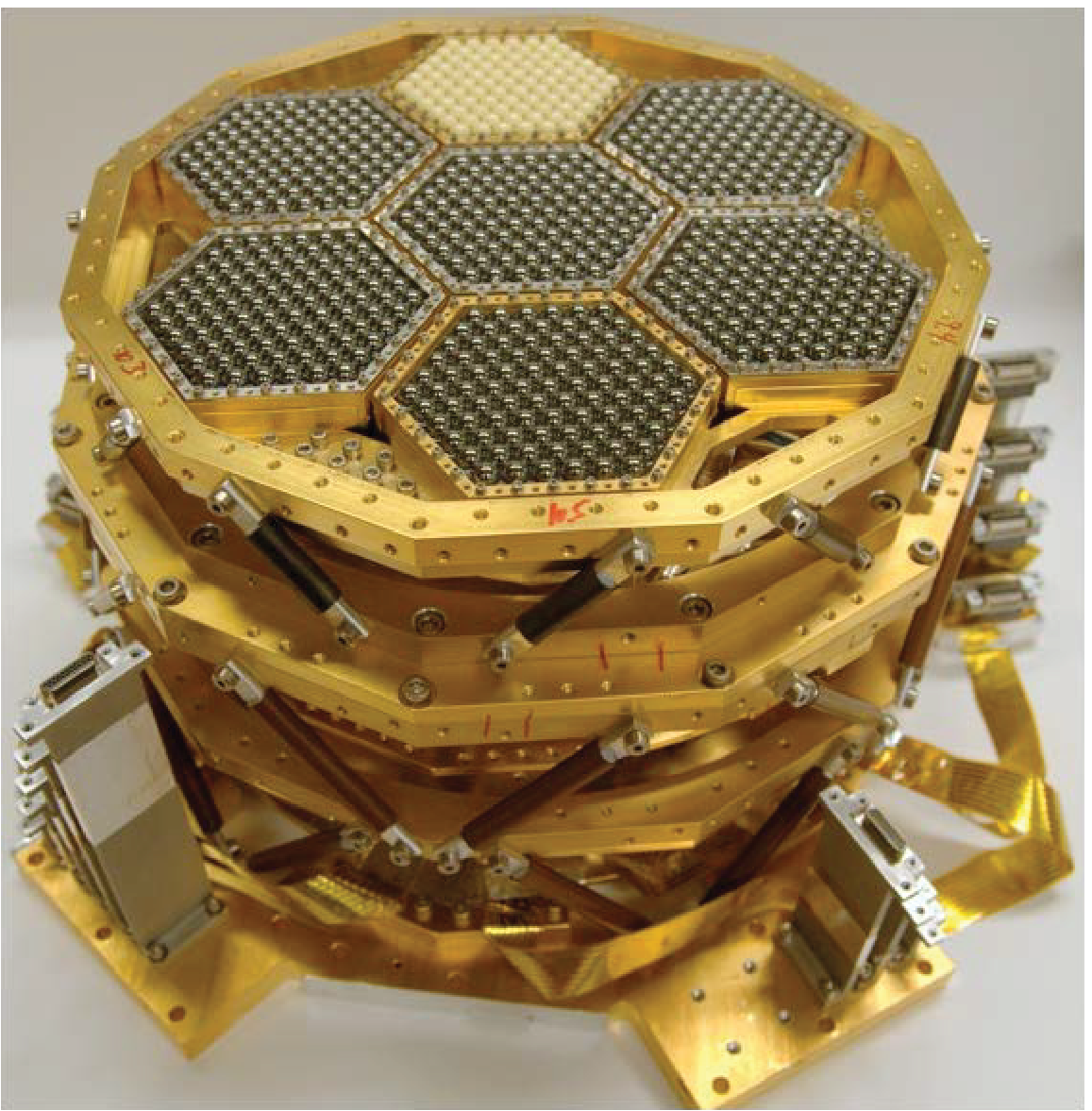} & \includegraphics[width=2.4in]{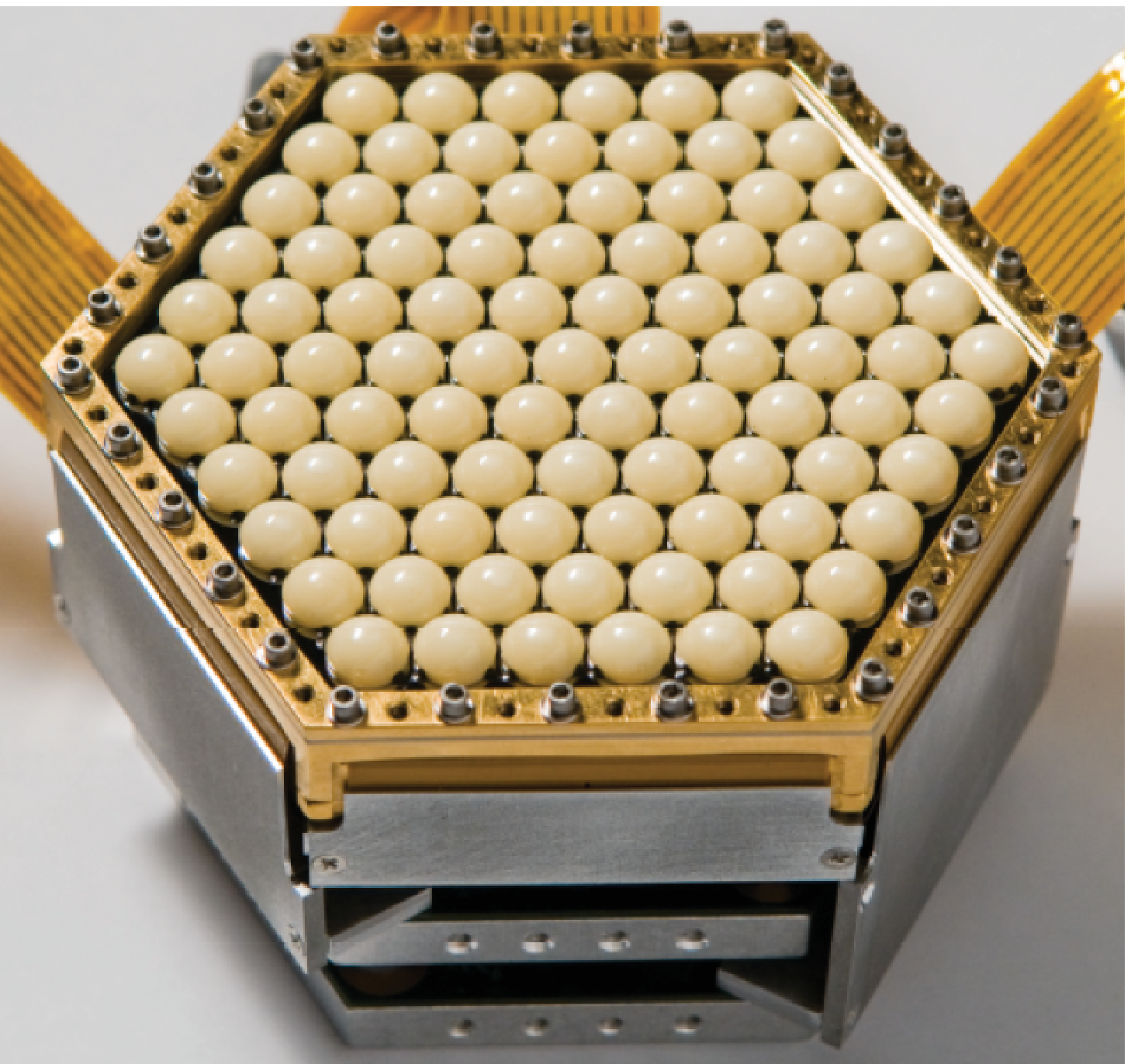} \\
\textbf{(a)} & \textbf{(b)}
\end{tabular}
   \end{center}
   \caption[Array]{\label{fig:arraypicture} \textbf{(a)} A picture of the entire assembled \pb\ focal plane. The 6 silver sub-arrays use silicon lenslets with thermoformed \acf{pei} \acf{ar} coatings, while the single white array uses alumina lenslets, a material with similar optical properties to silicon. \textbf{(b)} A picture of a single sub-array. Note how the readout folds back behind the sub-array for a compact, scalable design.}
   \end{figure} 

The \pb\ array consists of 637 pixels (1274 antenna-coupled bolometers) on 7 hexagonal sub-arrays. Each of these sub-arrays is fabricated on a single 4'' silicon wafer in the Berkeley Nanolab. Pictures of the assembled array and a single assembled sub-array are shown in \figref \ref{fig:arraypicture}.

\begin{figure}
   \begin{center}
\begin{tabular}{ccc}
\includegraphics[width=2in]{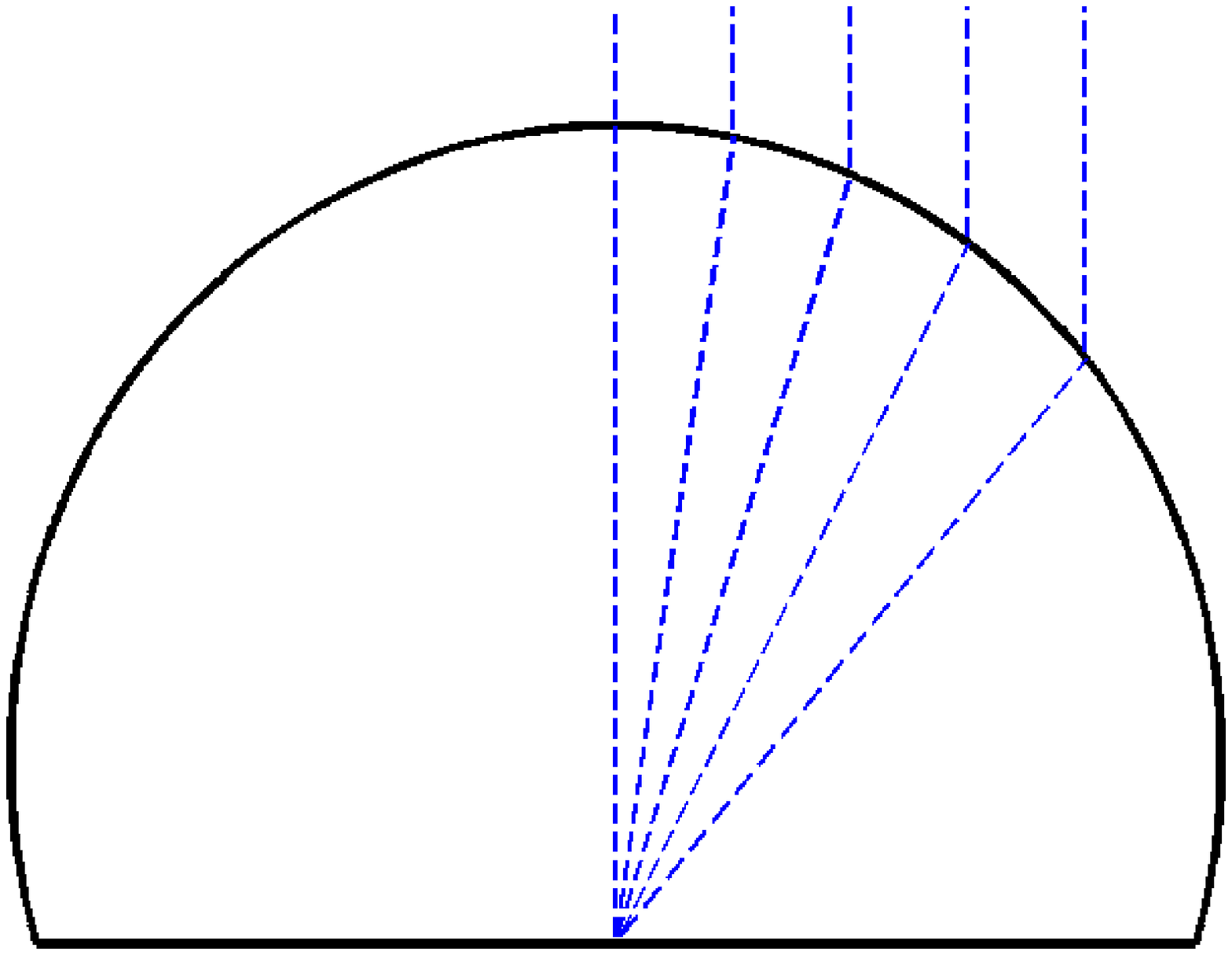} & \includegraphics[width=2.0in]{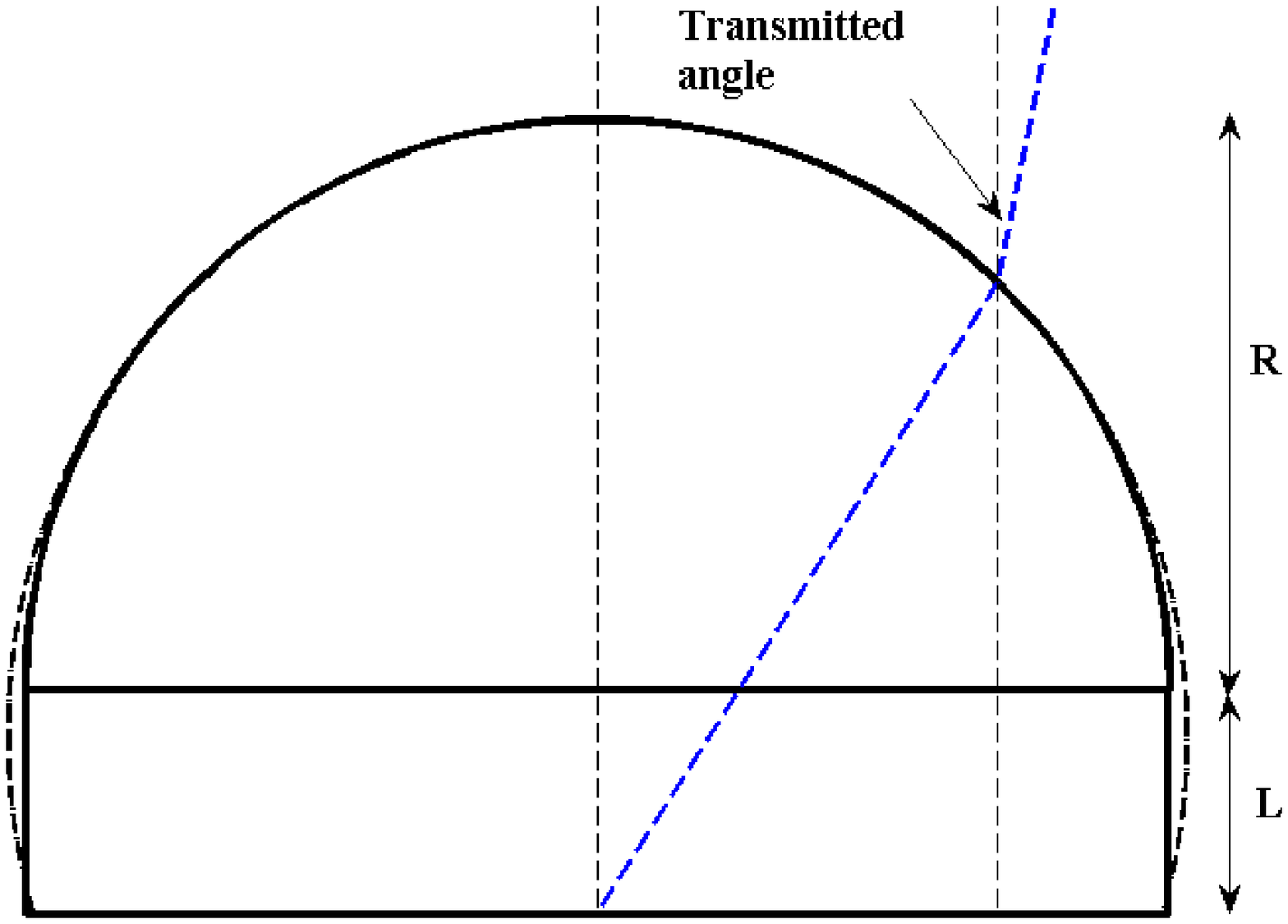} & \includegraphics[width=2.0in]{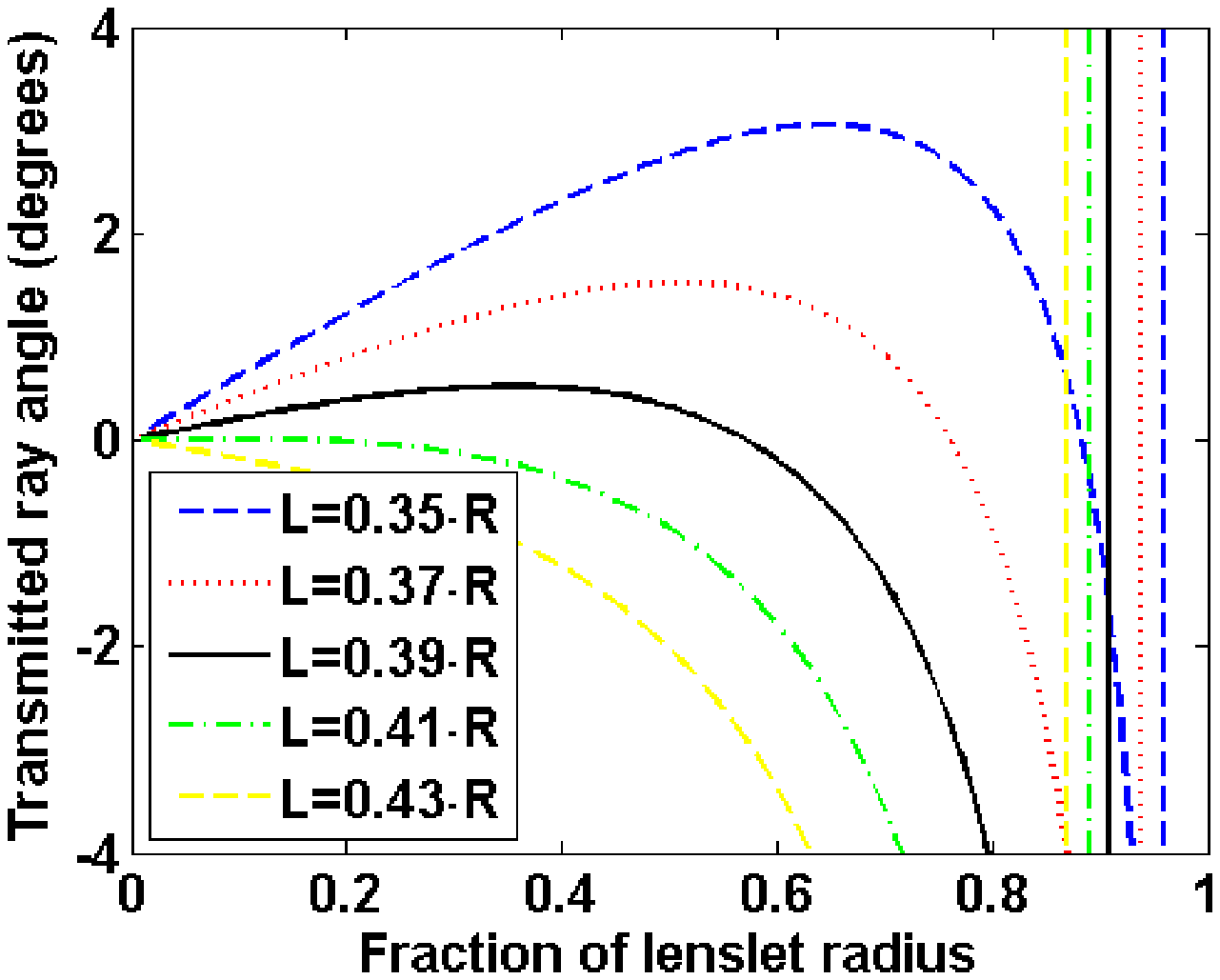} \\
\textbf{(a)} & \textbf{(b)} & \textbf{(c)}
\end{tabular}
   \end{center}
   \caption[Refraction at the surface of directly contacting dielectric lenslets]{\label{fig:ellipticalLenslet}Refraction at the surface of a dielectric lenslet for {\bf(a)} an elliptical lenslet and {\bf(b)} an extended hemisphere approximating the elliptical lenslet shape. In the case of the elliptical lenslet, all rays emanating from the center exit the lenslet vertically. In the case of the extended hemisphere, the transmitted angle depends on the extension length and the position on the lenslet surface. \textbf{(c)} Geometric calculation of the transmitted ray angle when exiting the lenslet surface for rays emanating from the center of the planar crossed double-slot dipole. For a true ellipse, this angle would be zero at all points along the radius of the lenslet. The five different lines are for different extension lengths (L) related to the pixel radius as shown in the legend. The vertical lines mark the point after which total internal reflection occurs.}
   \end{figure}

\begin{figure}
   \begin{center}
	\begin{tabular}{ccc}
\includegraphics[width=1.5in]{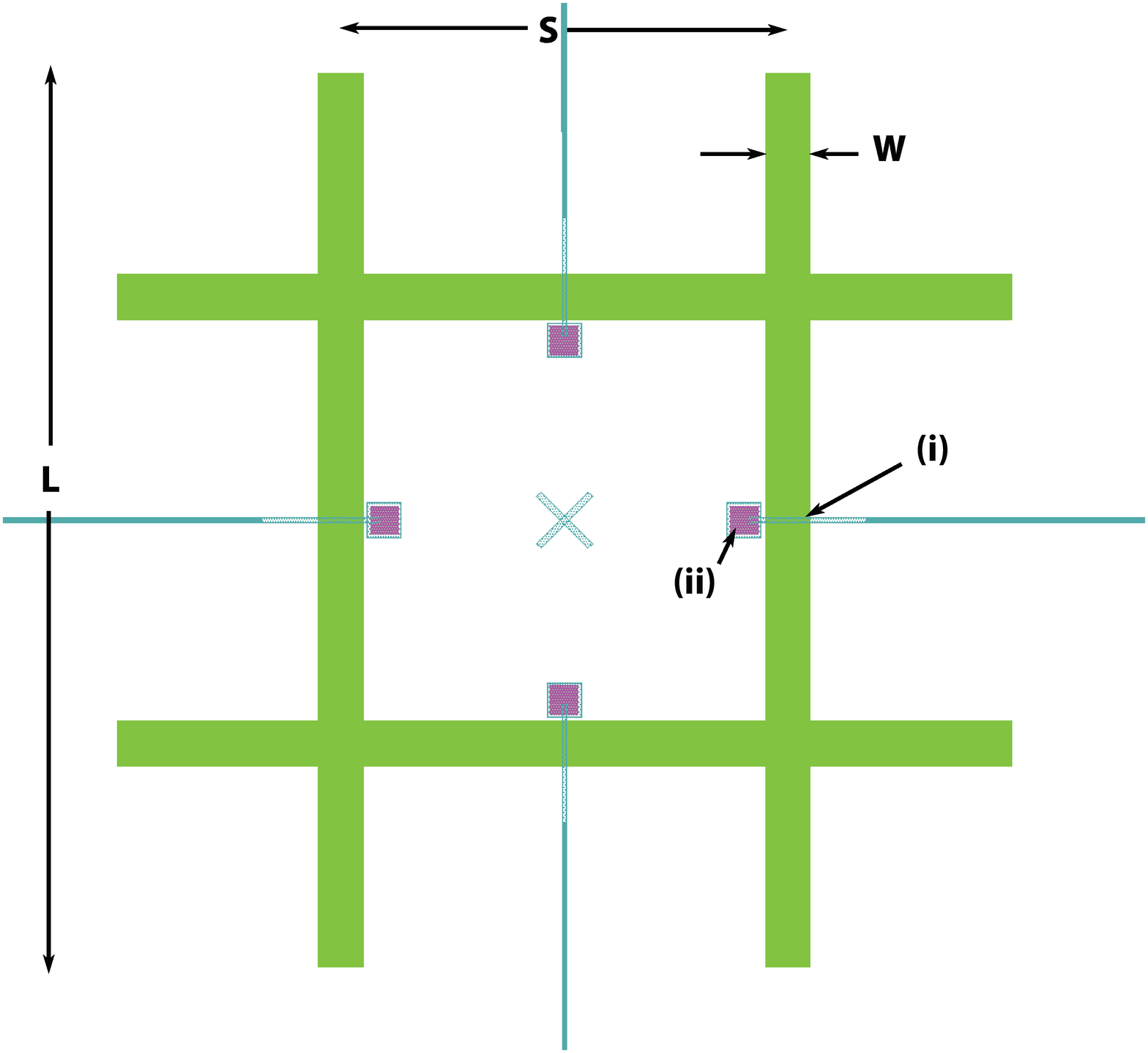} & \includegraphics[width=2.5in]{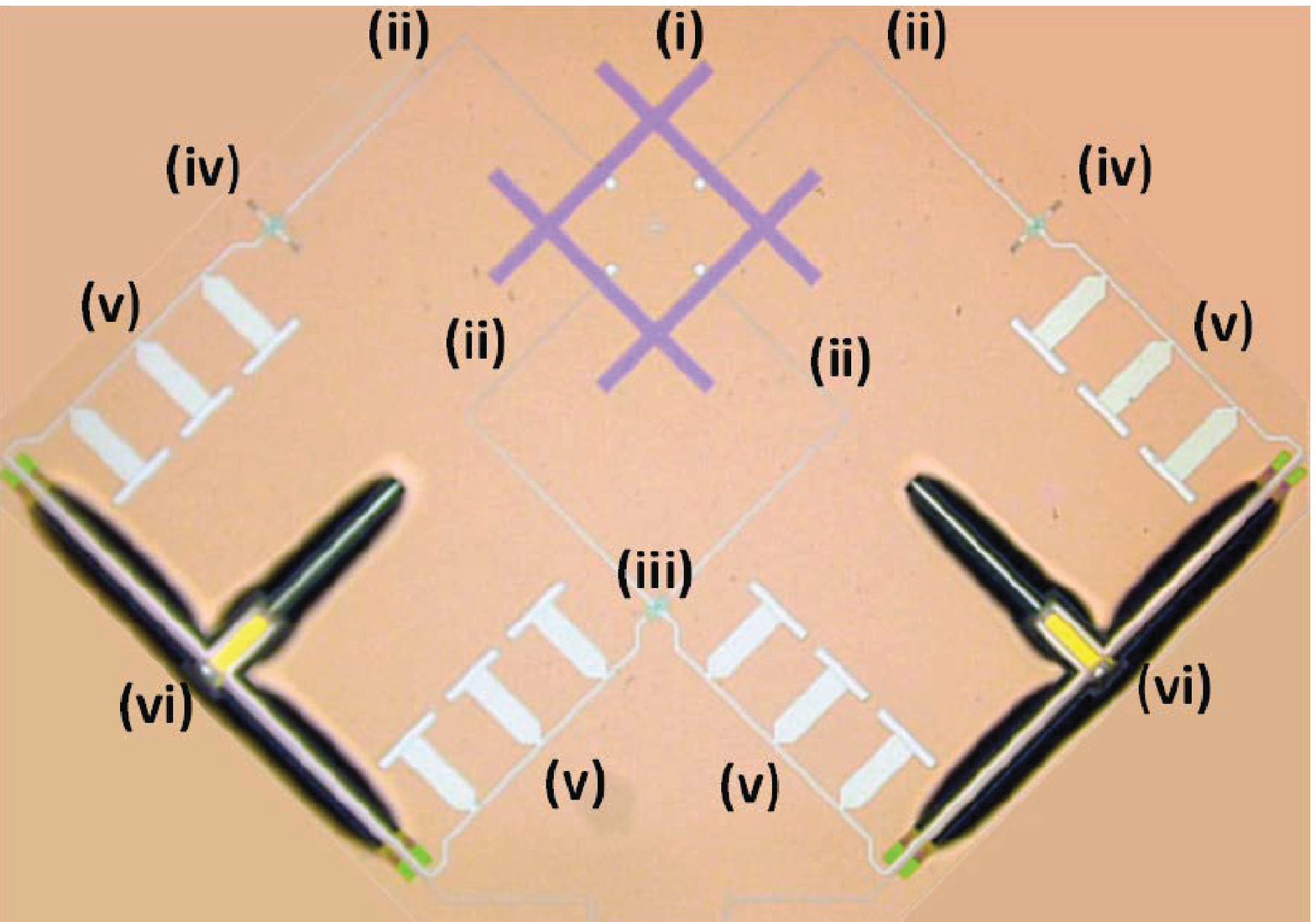} & \includegraphics[width=2.2in]{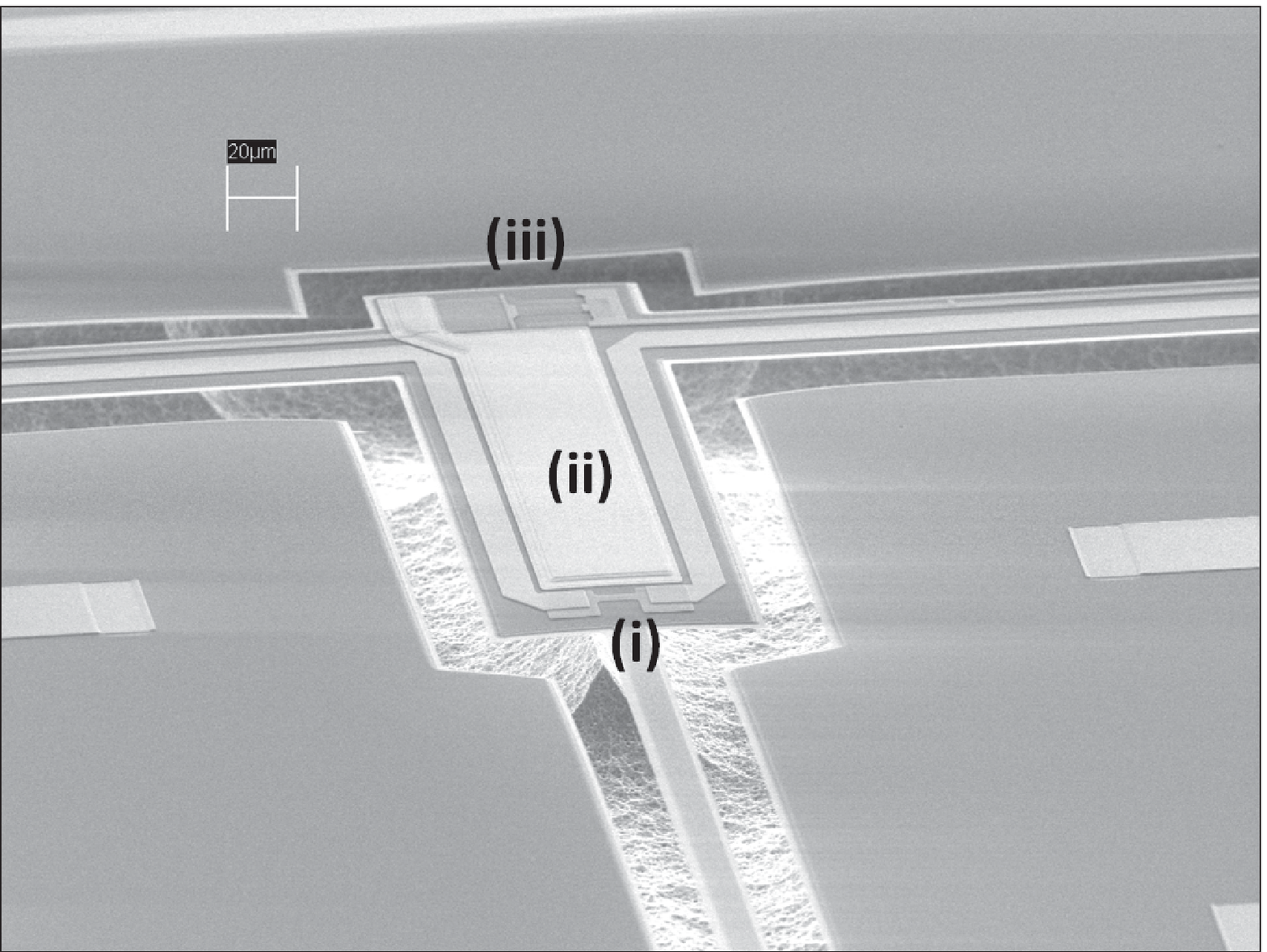} \\
\textbf{(a)} & \textbf{(b)} & \textbf{(c)}
\end{tabular}
   \end{center}
   \caption[Pixel photograph]{\label{fig:pixel}\textbf{(a)} Design drawing of the planar crossed double-slot dipole. Areas in green represent slots in the niobium ground plane. \textbf{(i)} Microstrip feeds are shown at the center of each rectangular slot, with a \textbf{(ii)} via from the microstrip to the ground plane (in purple) internal to each slot. The incoming radiation excites a voltage across the slot and current around the slot, which excites a propagating wave in the microstrip. For the \pb\ pixel, $W=33.9~\mu$m ($0.017\lambda_0$), $S=338.7~\mu$m ($0.167\lambda_0$), and $L=677.5~\mu$m ($0.334\lambda_0$). \textbf{(b)} A photograph of the planar structures in one of the \pb\ pixels. The structures are labelled in the order that the \ac{cmb} photon travels through them: {\bf(i)} Crossed double-slot dipole, {\bf(ii)} Dolph-Chebychev microstrip transformer {\bf(iii)} microstrip cross-under, {\bf(iv)} cross-under balancing structures, {\bf(v)} microstrip filters, and {\bf(vi)} bolometers. \textbf{(c)} Scanning electron micrograph of a \pb\ bolometer, showing \textbf{(i)} the resistive load onto which the observed optical power is dissipated, \textbf{(ii)} the dual-transition \ac{tes} that senses modulation of this power, and \textbf{(iii)} the gold added to the bolometer to increase its heat capacity.}
   \end{figure}

The \pb\ detectors integrate several features into a single pixel: polarization-sensitive antennas separate the signal into polarized components and couple these components from free space into superconducting microstrip waveguide, spectral bandpass filtering transmits only the desired frequency band, and superconducting \ac{tes} thermistors on the thermally released bolometers provide the photon-noise limited detection of the transmitted electromagnetic wave. An overview of this architecture is shown in Figures \ref{fig:ellipticalLenslet} and \ref{fig:pixel}. 

\subsection{Antenna design: lenslet-coupled crossed double-slot dipole}
\label{sec:antennaDesign}

The idea of increasing an antenna's directivity with a directly contacting dielectric lenslet has been used across many applications and spectral regions \cite{RutledgeFirstDielLens,Zmuidzinas_quasi-optical,Filipovic_double_slot,LensletReferenceMap}. This configuration has several benefits:

\begin{itemize}

\item Because there is a difference in near-field dielectric constant between the sides of the planar antenna, the beam is formed more strongly in the direction of the high dielectric constant. In the case of silicon, with a relative dielectric constant of 11.7, the antenna gain is 10 times higher in the forward direction (the direction of the lenslet) than in the backward direction\cite{Filipovic_double_slot}.

\item Standard lithographic techniques produce planar antennas on dielectric substrates. While the gain is preferentially in the direction of the substrate as described above, total internal reflection at the substrate boundry leads the antenna to be sensitive to modes propagating within the planar substrate. An extended hemisphere of dielectric material limits that total internal reflection.

\item The hemispherical dielectric lenslet acts as a beam-forming element, coupling the planar structure's broad beam within the dielectric to a more directed beam in free space---the lenslet acts to magnify the active size of the antenna. This leaves room under the lenslet (away from its center) for other microwave elements, without using the valuable focal plane area coupled to the diffraction-limited field of view of the telescope.

\item Refracting optics have a wide bandwidth of operation. Given that the lenslet medium has low loss, the loss in coupling to free space is dominated by the reflection loss at the surface of the lenslet. This can be mitigated using an \ac{ar} coating. For the \pb\ pixel, a single-layer quarter-wavelength \ac{ar} coating is sufficient to reduce the reflection loss over the spectral band of interest. In the future, multichroic antennas coupled to lenslets with multi-layer antireflection coatings will have the capability to measure several spectral bands through a single pixel \cite{Suzuki_SPIE2012,OBrient:Thesis,OBrient2008SPIE}.

\end{itemize}

In the geometric limit, rays from a point source emitter within a dielectric will refract to be parallel if the surface of the dielectric is the far side of an ellipse of the correct eccentricity \cite{Hecht}. \figref \ref{fig:ellipticalLenslet}{\bf(a)} shows this refraction for an ellipse of the eccentricity required given the optical index of silicon. For lenslets of high dielectric constant this eccentricity is small, and thus the ellipse can be well-approximated by an extended hemisphere. Figure \ref{fig:ellipticalLenslet}{\bf(b)} shows a cross section of such an extended hemispherical lenslet, and \figref \ref{fig:ellipticalLenslet}\textbf{(c)} shows the transmitted angle for such a lenslet as a function of position on the lenslet surface, for several extension lengths. These angles are calculated using geometric optics; to the extent the angles are close to zero, the electromagnetic wave couples to a diffraction-limited beam with waist diameter slightly smaller than the lenslet diameter, just as with a horn antenna. 

The nominal \pb\ pixel size is $R=1.6\lambda_0$, with $\lambda_0$ the free-space wavelength of the center of the spectral band. For a range of lenslet extension lengths $L$ between $0.36R$ and $0.39R$, the lenslet-coupled double-slot dipole antenna produces a beam with high Gaussisity and directivity \cite{Filipovic_double_slot}. The \pb\ dielectric lenslet extensions have some fabrication variability; the above values were chosen as bounds for acceptable extension thickness.   

For lenslets of high dielectric constant, the reflection loss at the surface of the lenslet is significant. This reflection loss can be mitigated by employing a quarter-wavelength \ac{ar} coating at the lenslet surface. As the extension length increases, the reflectance at the quarter-wavelength coating increases because of the non-normal incidence of the radiation and the curvature of the surface \cite{Filipovic_double_slot}. A re-optimization of the coating thickness as a function of distance from the center of the lenslet dictates that the coating be slightly thicker at the side of the lenslet, but this optimization was not attempted for the \pb\ lenslets, which were \ac{ar} coated with uniform-thickness thermoformed polyetherimide, a plastic of optical index 1.7.

The planar structure coupled to the dielectric lenslet must be able to couple photons efficiently into the on-wafer  waveguide across the desired spectral band, have low cross-polarized response across its spatial beam (at least within the angles that transmit through the aperture stop), and couple efficiently through the lenslet to the diffraction-limited telescope optics. The resonant crossed double-slot dipole used in \pb\ was adapted from the antenna presented in Chattopadhyay and Zmuidzinas \cite{doubleSlotAntennaLetter}. This antenna is useful for its relatively low drive impedance, low cross-polarized response, and circular beams \cite{doubleSlotAntennaTest,OBrient:Thesis}.


\subsection{Dolph-Chebychev microstrip transformer}

\begin{figure}[htpb]
\begin{center}
\includegraphics[width=2.5in]{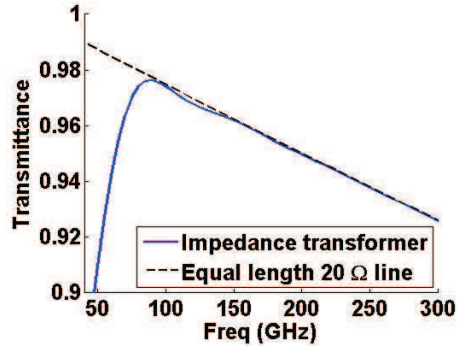}
\end{center}
\caption[Transmittance of the microstrip transformer]{\label{fig:xfmr}Transmitted power ($\left | S_{21} \right | ^2$) as a function of frequency for the microstrip transformer that connects the antenna to the microstrip filter. Also plotted is the transmittance due only to dielectric loss in a line of equal length.}
\end{figure}

The antenna described above has a resonant impedance of $30~\Omega$ \cite{doubleSlotAntennaLetter,OBrient:Thesis}. This impedance is achievable with the microstrip waveguide used to feed the antenna, but because of limitations in the resolution and repeatability of the optical lithography used to print the microstrip, wider (lower impedance) microstrip can be fabricated with greater repeatability. For this reason, the band pass filter is designed to couple to a lower input and output impedance of $10~\Omega$. A microstrip transformer is used to transition from the antenna impedance to the filter input impedance. Four of these microstrip transformers are used in each pixel, as shown in \figref \ref{fig:pixel}. As a function of the length along the transformer, $l$, the design microstrip impedance, $Z$, is given by \cite{transformerApproxAndTest}

\begin{equation}
Z(l) = Z_{antenna} \cdot e^{ \frac{1}{2} \ln \left ( \frac{Z_{filter}}{Z_{antenna}}\right ) \left [\sin \left (\pi\left ( \frac{l}{L}-\frac{1}{2} \right ) \right )+1 \right ] },
\end{equation}

\noindent where $L$ is the total length of the transformer. The transmittance of that transformer, including the effect of the dielectric loss in the microstrip, is shown in \figref \ref{fig:xfmr}.

\subsection{Microstrip cross-under}
\label{sec:cross-under}

\begin{figure}[htpb]
\begin{center}
\begin{tabular}{cc}
\includegraphics[width=2.0in]{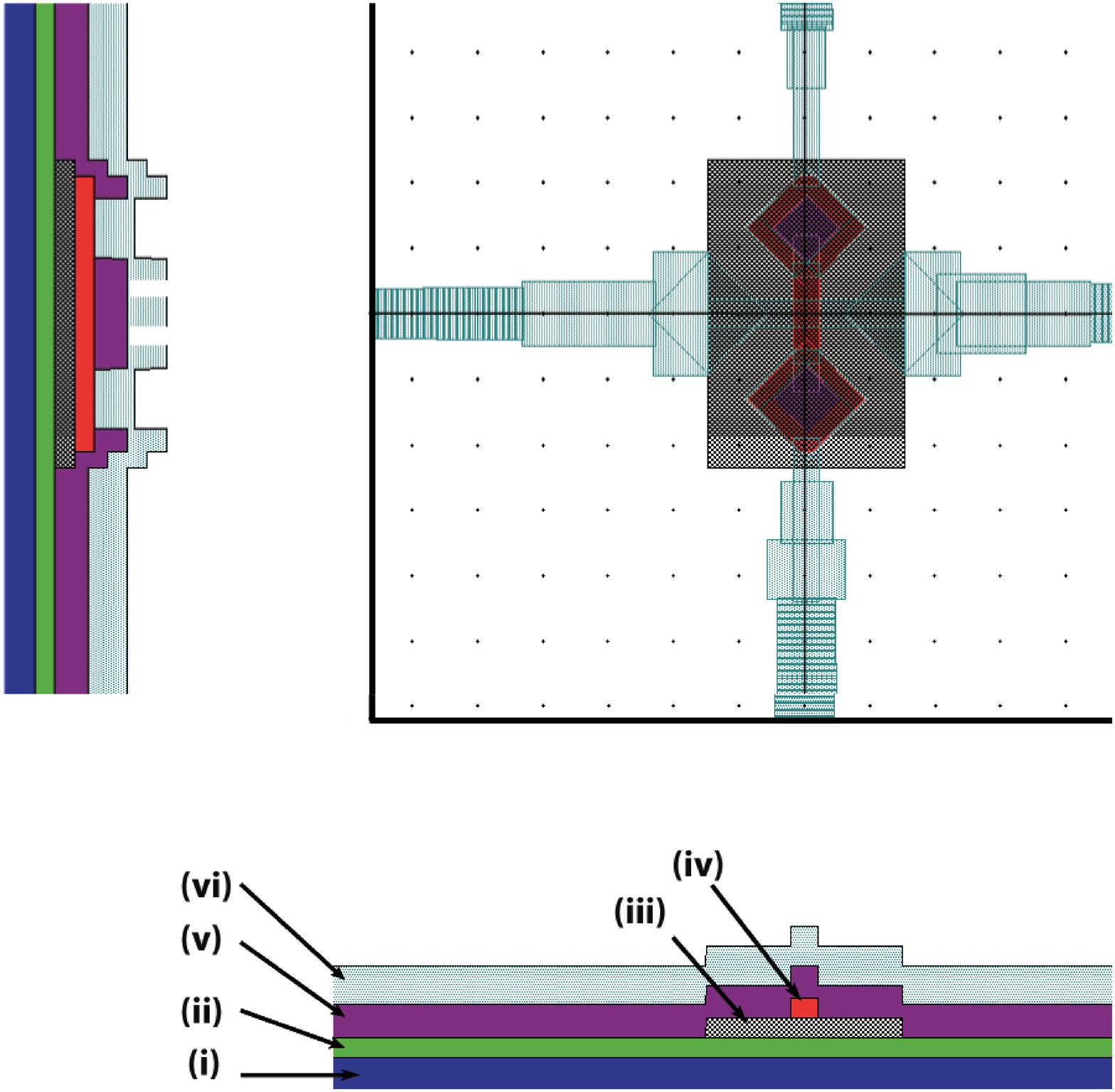} & \includegraphics[width=1.8in]{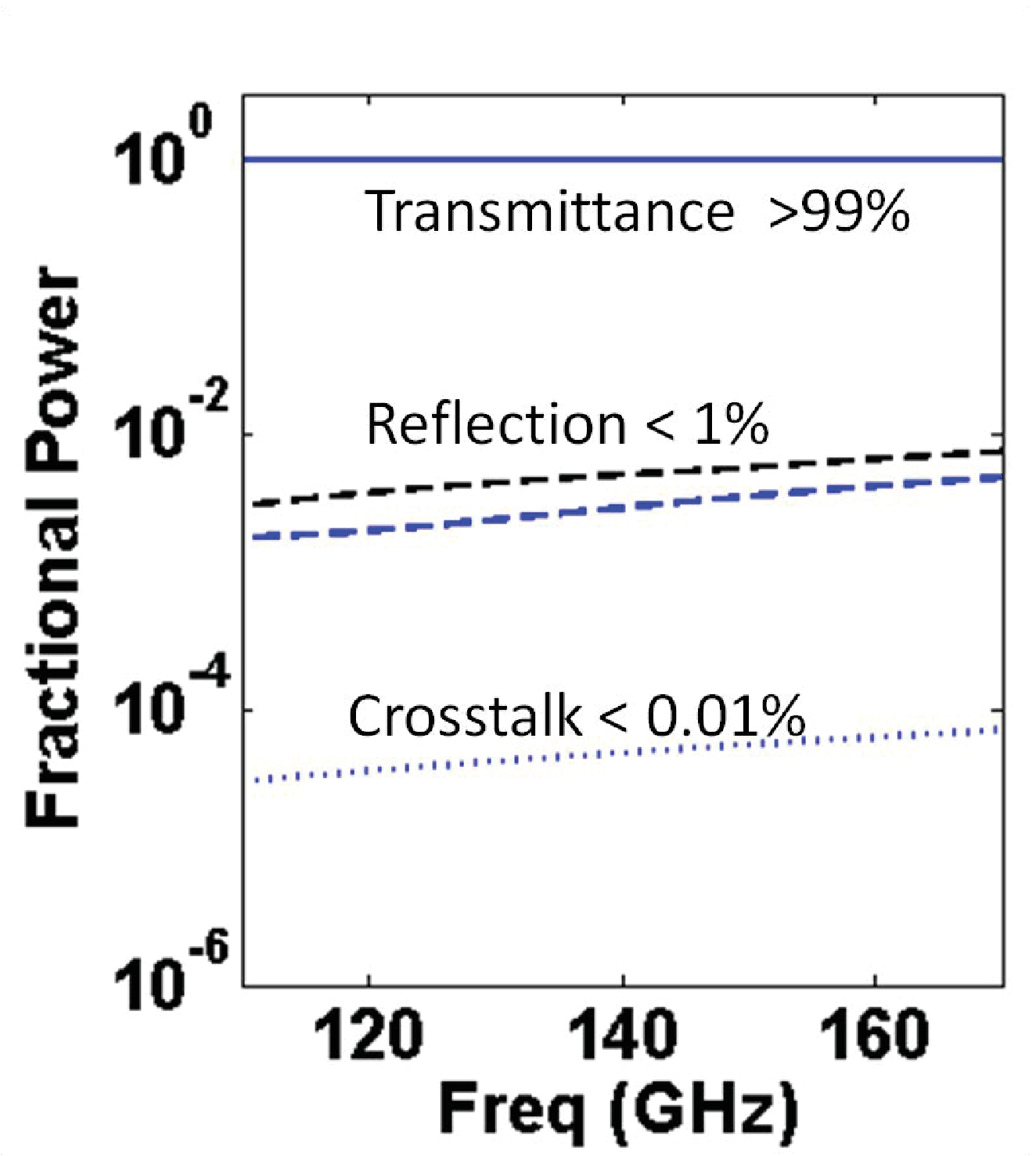} \\
{\bf(a)} & {\bf(b)}
\end{tabular}
\end{center}
\caption[Microstrip cross-under design and simulated performance]{\label{fig:crosssim}Microstrip cross-under {\bf(a)} design and {\bf(b)} simulated performance. {\bf(a)} shows cross-section views from two angles. In the cross-sections, the vertical scale is enhanced by a factor of 10 for clarity. Layers shown are \textbf{(i)} Low-stress silicon nitride, \textbf{(ii)} Nb ground plane, \textbf{(iii)} SiO$_2$ crossunder isolation, \textbf{(iv)} Nb cross-under, \textbf{(v)} SiO$_2$ dielectric, and \textbf{(vi)} Nb microstrip conductor. {\bf(b)} Transmittance, reflection, and crosstalk as simulated in Sonnet.}
\end{figure}

Because of the topology of the antenna, the signals in two of the microstrip lines must cross each other. This is accomplished using a microstrip cross-under structure. The design of this structure is shown in \figref \ref{fig:crosssim}{\bf(a)}. The cross-under was simulated using the Sonnet full-wave simulation package\footnote{http://www.sonnetsoftware.com/}. The results of this simulation are shown in figure \ref{fig:crosssim}{\bf(b)}. The reflected power is less than 1\% and the crosstalk power is less than 0.01\%. 

\subsection{Bandpass filter}

The bandpass filter is a standard resonant filter design: a shorted quarter-wavelength stub filter \cite{MstripFilterDesignEquations}. Note that the stubs are narrowed where they meet the trunkline. This is necessary for the phase fronts of the wave on the low impedance stub to be perpendicular to the stub length. When the superconducting niobium microstrip is correctly modeled \cite{KerrSonnetCorrection}, an analytical network model of the filter closely reflects the results of a full-wave simulation using Sonnet. Using the analytical model, the parameters of the microstrip filter---center frequency, bandwidth, and steepness of edge (number of poles)---were optimized given the expected atmospheric emission and absorption profile.

\subsection{Bolometer structure}

A bolometer is fundamentally a place where radiative power is deposited that is in thermal contact with a thermistor of heat capacity $C$, all of which has a weak connection to a thermal bath.

In the case of \pb, the weak thermal link between the thermistor and the thermal bath is provided by a silicon nitride suspension, where the silicon substrate is etched away beneath the suspension. The scanning electron micrograph in \figref \ref{fig:pixel}\textbf{(c)} shows this suspension. The thermal path between the thermistor and the silicon wafer is along the silicon nitride suspension, which is $\sim 1~\mu$m in thickness. The operating power of the bolometer is set by the thermal conductance of this suspension and the thermistor's superconducting transition temperature.

The electrothermal bandwidth of the bolometer $\Delta f_{et}$ is proportional to its small-signal thermal conductance, $g$, the electrothermal loop gain $\mathcal{L}$, and inversely proportional to its heat capacity, $C$ \cite{Gildemeister:Thesis}. Whatever the electrothermal bandwidth, the readout bandwidth for that bolometer must be larger by a factor of $\sim 5$ to stably voltage bias the bolometer\cite{Irwin:etbw}.  The scanning speed of the telescope limits the bandwidth over which there is useful astrophysical signal. The limited readout bandwidth available can best be used multiplexing several bolometers, each with only the necessary electrothermal bandwidth, rather than fewer bolometers of larger $\Delta f_{et}$. Since $g$ is set by the bolometer properties and the desired operating power, $C$ is controlled in fabrication to achieve the desired $\Delta f_{et}$ at reasonable $\mathcal{L}$. 

The head capacity, $C$, is controlled by depositing gold on the bolometer and making sure the gold is in good thermal contact with the thermistor \cite{JaredThesis}. \figref \ref{fig:pixel}\textbf{(c)} shows the gold used for this purpose. The thermistor material extends underneath the gold, and the gold extends onto the thermistor leads, so as to make the best thermal contact possible without strongly affecting current flow across the thermistor.  

\subsubsection{Resistive load}

The optimal resistive load for absorbing power from two balanced $10~\Omega$ microstrip lines has a real impedance of $20~\Omega$ and no imaginary impedance, creating a virtual ground in the center of the load resistor. The \pb\ load resistor is fabricated from titanium with a sheet resistance $\sim 10~\Omega/sq$. It's length was adjusted to achieve a measured DC resistance of $20~\Omega$.

\subsubsection{Dual-TES thermistors}
\label{sec:dualtes}

The \pb\ bolometers employ a dual-\ac{tes} thermistor \cite{DualTES} to allow for two modes of observation: a low noise mode for cosmological observations and a higher noise mode with larger dynamic range. The low noise mode uses an aluminum/titanium bilayer thermistor with a transition temperature of $T_{c,bilayer} \approx 0.5$ K, while the mode with larger dynamic range employs a bare aluminum thermistor with a transition temperature of $T_{c,Al} = 1.4$ K. The design of the bilayer thermistor is somewhat independent of the Al thermistor. When operating the bolometer with the bilayer thermistor, the Al thermistor is simply a part of the superconducting thermistor leads. The bilayer thermistor operating resistance is $\sim 1~\Omega$, which provides the bolometer responsivity necessary and couples correctly to the readout system.

\section{Focal plane array fabrication}
\label{sec:fab}

A single sub-array consists of a hexagonal device wafer of antenna-coupled bolometers on one side, with its other side in contact with a hexagonal array of lenslets. This section describes the fabrication of both of these items. 

\subsection{Lenslet sub-array}

The hemispherical lenslets are produced from semiconductor-grade single-crystal silicon. They are ground into spheres and then ground flat along a plane intersecting a great circle of the sphere\footnote{The commercial grinding process is done by Rayotek Scientific, Inc.}. This leaves a hemisphere to be integrated with a flat silicon spacer and the antenna as described in \secref \ref{sec:antennaDesign}.

The lenslet sub-array acts both as part of the extension length providing the synthesized elliptical lenslet shape, and as an alignment wafer to properly center each lenslet on its associated crossed double-slot dipole. Circles of diameter slightly ($\sim 40~\mu$m) larger than the diameter of the lenslets are etched into the lenslet wafer using a deep reactive ion etching process \cite{Bosch}. These precise circular depressions serve to locate each lenslet to within $\sim \lambda_0/100$. The entire lenslet wafer is then optically aligned to the device wafer using transmission infrared microscopy.

\subsection{Antenna-coupled bolometer sub-array}

   \begin{table}[htpb]
   \begin{center}
	\begin{tabular}{|c|c|c|c|}
\hline
\textbf{Material} & \textbf{\shortstack{Thickness\\(microns)}} &\textbf{Deposition Process} & \textbf{Etch Process} \\
\hline
Silicon substrate & 500 & Prime-grade wafers & XeF$_2$ vapor \\
Silicon dioxide & 0.05 & Thermal growth & CF$_4$ RIE \\
Low stress nitride & 0.7 & LPCVD & CF$_4$ RIE \\
Niobium & 0.3 & DC magnetron sputter & CF$_4$ / O$_2$ RIE \\
Silicon dioxide & 0.3 & 350\textdegree C PECVD & CHF$_3$ / O$_2$ RIE \\
Niobium & 0.3 & DC magnetron sputter & CF$_4$ / O$_2$ RIE \\
Silicon dioxide & 0.5 & 350\textdegree C PECVD & CHF$_3$ / O$_2$ RIE \\
Niobium & 0.6 & DC magnetron sputter & CF$_4$ / O$_2$ RIE \\
Aluminum/titanium bilayer & 0.04/0.12 & DC magnetron sputter & Premixed wet etch / SF$_6$ RIE \\
Gold & 1.5 & Evaporation & Liftoff in acetone \\
\hline

	\end{tabular}
	\end{center}
	\caption[Layers of the \pb\ device wafer]{\label{tab:process}Layers in the \pb\ device wafer, along with thickness, deposition process, and etching process. Acronyms used in the table: RIE: Reactive Ion Etch; CVD: Chemical Vapor Deposition; PECVD: Plasma-enhanced CVD; LPCVD: Low-Pressure CVD. }
	\end{table}

The \pb\ device wafers require an eight layer process on a silicon substrate. The process is outlined in \tabref \ref{tab:process}, in order of layer deposition. The etching process for each layer is given; the etches for the layer are not in order, because they do not all occur immediately after deposition.

\subsubsection{Substrate}

The silicon substrate forms part of the dielectric contacting lenslet, so its microwave loss properties are important. A \ac{fts} was used to measure loss in commercial semiconductor-grade silicon at the frequencies of interest. The upper bound for loss on this measurement was low enough that it was not a significant source of loss in the pixel. It is important to note that some silicon with a higher impurity content did have significant loss and was unacceptable for this application.

\subsubsection{Photoresist coating and patterning}

Patterning is done in standard positive photoresist with a thickness of 1--4 microns, where the exact recipe depends on the process step. Patterning of the individual pixels is accomplished using a step and repeat 10X optical reduction printer, a GCA wafer stepper. Test pixels, wafer border, and alignment marks are also printed by the wafer stepper. The niobium thermistor leads, which transmit signals from the edge of the wafer to the thermistors, are patterned using a wafer-scale contact printer. After each patterning step, the photoresist undergoes a post-exposure bake and then an automated puddle develop. After development, the wafer is exposed to a 1 minute low-power oxygen plasma to remove any residual photoresist where the pattern was developed.

\subsubsection{Microwave structures}

The continuous niobium ground plane is broken only where necessary for the bolometer release and for the slots of the crossed double-slot dipole. Those slots are connected to superconducting microstrip waveguides that consist of a niobium ground plane, silicon dioxide dielectric, and niobium top conductor. The dielectric loss in the silicon dioxide is the primary source of loss in this superconducting microstrip transmission line \cite{Myers:Thesis}.

The microwave structures, including the cross-under, are constructed of three layers of niobium and two layers of silicon dioxide. Note that the silicon dioxide is not deposited in a furnace. This is important because high temperatures degrade the quality of the niobium as a superconductor \cite{NiobiumOxideTc}. Also note that the thicknesses of the niobium ground plane and microstrip are 300 and 600 nm, respectively. At this thickness, both films are significantly thicker than the penetration depth \cite{NiobiumPenetrationDepth}, which reduces the dependence of the microstrip impedance on film thickness \cite{YassinWithingtonMicrostrip}.

\subsubsection{Bolometer structure}

The \ac{tes} thermistor is an aluminum/titanium bilayer. The layers of this bilayer must be deposited in direct succession without the wafer leaving the vacuum environment, and with an etch before the deposition to remove the native niobium oxide on the niobium thermistor leads. The gold used to increase the heat capacity of the bolometer is deposited using electron beam evaporation onto already patterned photoresist, so that the gold is removed everywhere other than where it is desired by dissolving the photoresist in a lift-off procedure. The \ac{tes} bolometer suspension structure is provided by the Low-stress Silicon Nitride (LSN) film deposited directly on the silicon wafer. This deposition is first, but the etching of the LSN occurs much later in the process, after all of the deposition steps are complete. To release the suspended \ac{tes} structure, trenches are etched into the LSN, and then the silicon substrate underneath the LSN is removed by an isotropic gaseous xenon difluoride ($\mbox{XeF}_2$) etch. The rough surface below the released bolometer (visible in \figref \ref{fig:pixel}\textbf{(c)}) is where the silicon has been etched by the $\mbox{XeF}_2$, leaving the suspended bolometer structure.

\section{FOCAL PLANE PERFORMANCE VALIDATION}
\label{sec:valid}

This section reports on several of the tests that were conducted to validate the \pb\ detector performance. 

\subsection{Pixel beam maps}

\begin{figure}[htpb]
   \begin{center}
	\begin{tabular}{cc}
   \includegraphics[width=2.5in]{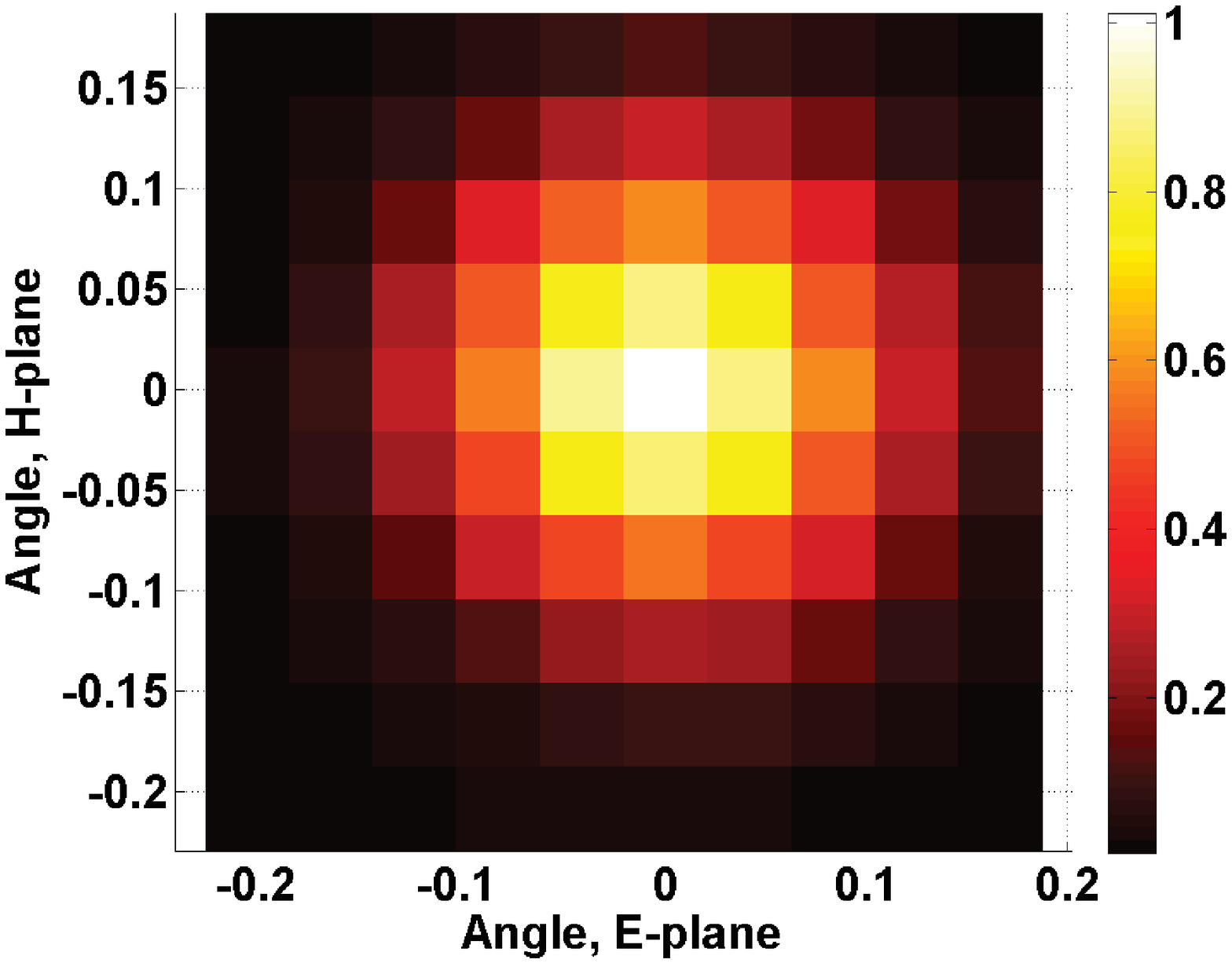} & \includegraphics[width=2.5in]{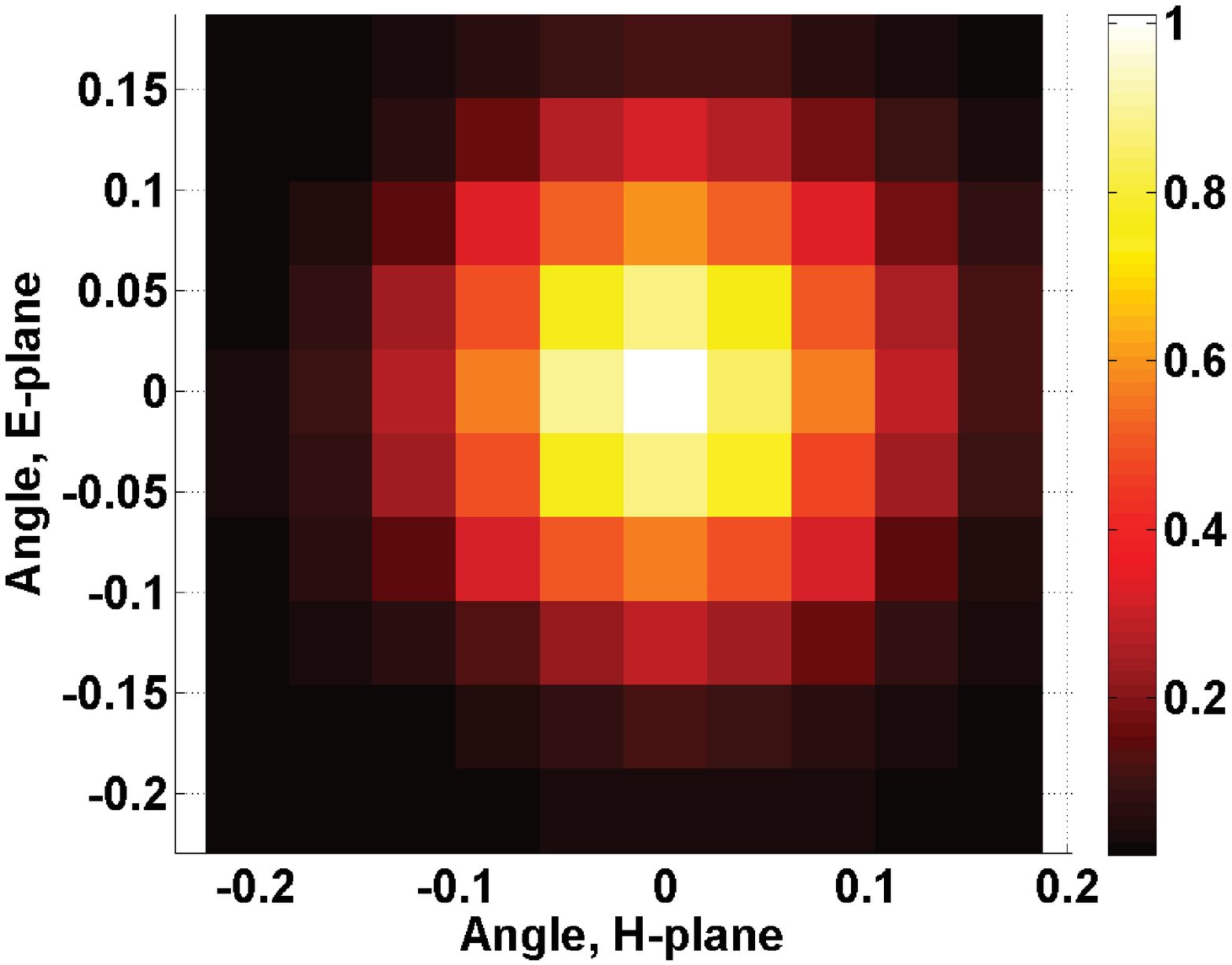} \\
{\bf(a)} & {\bf(b)}
\end{tabular}
	\end{center}
	\caption[Pixel beam maps]{\label{fig:beammaps}Spatial beam map of the \pb\ pixel response for each polarization, as measured using a chopped thermal source on a translation stage.}
	\end{figure}

Maps of the detectors' spatial response patterns were measured in the laboratory using a modulated thermal source on translation stages. Figure \ref{fig:beammaps} shows the result of that measurement for each polarization of a single pixel. Note that the beam ellipticity is small (best fit elliptical Gaussian beams to this data have ellipticities of 1\%), and that the differential pointing is small, with a best fit value of 0.3\% of the FWHM.

\subsection{Boresight cross-polar response}

\begin{figure}[htpb]
   \begin{center}
	\begin{tabular}{cc}
   \includegraphics[width=2.9in]{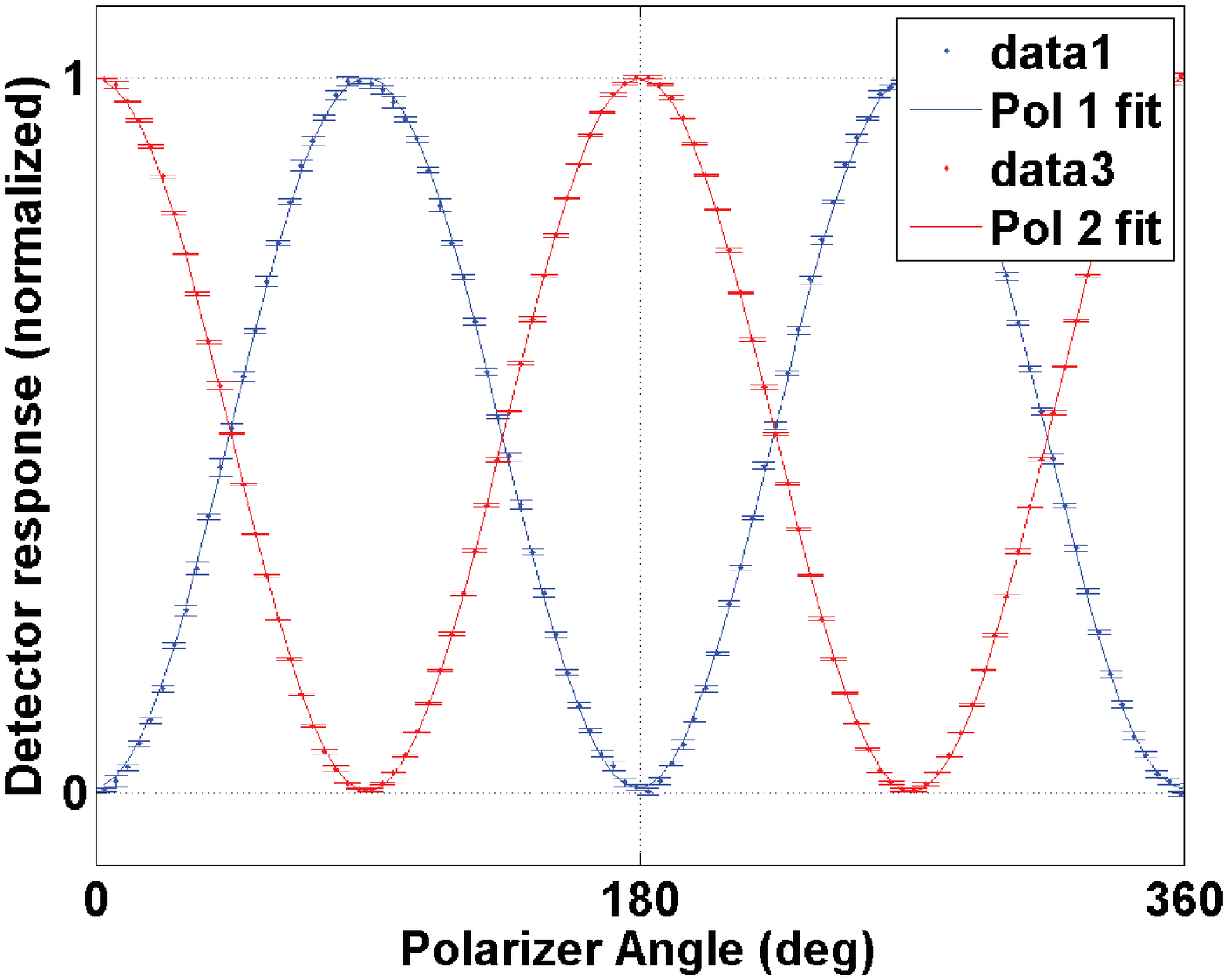} & \includegraphics[width=3in]{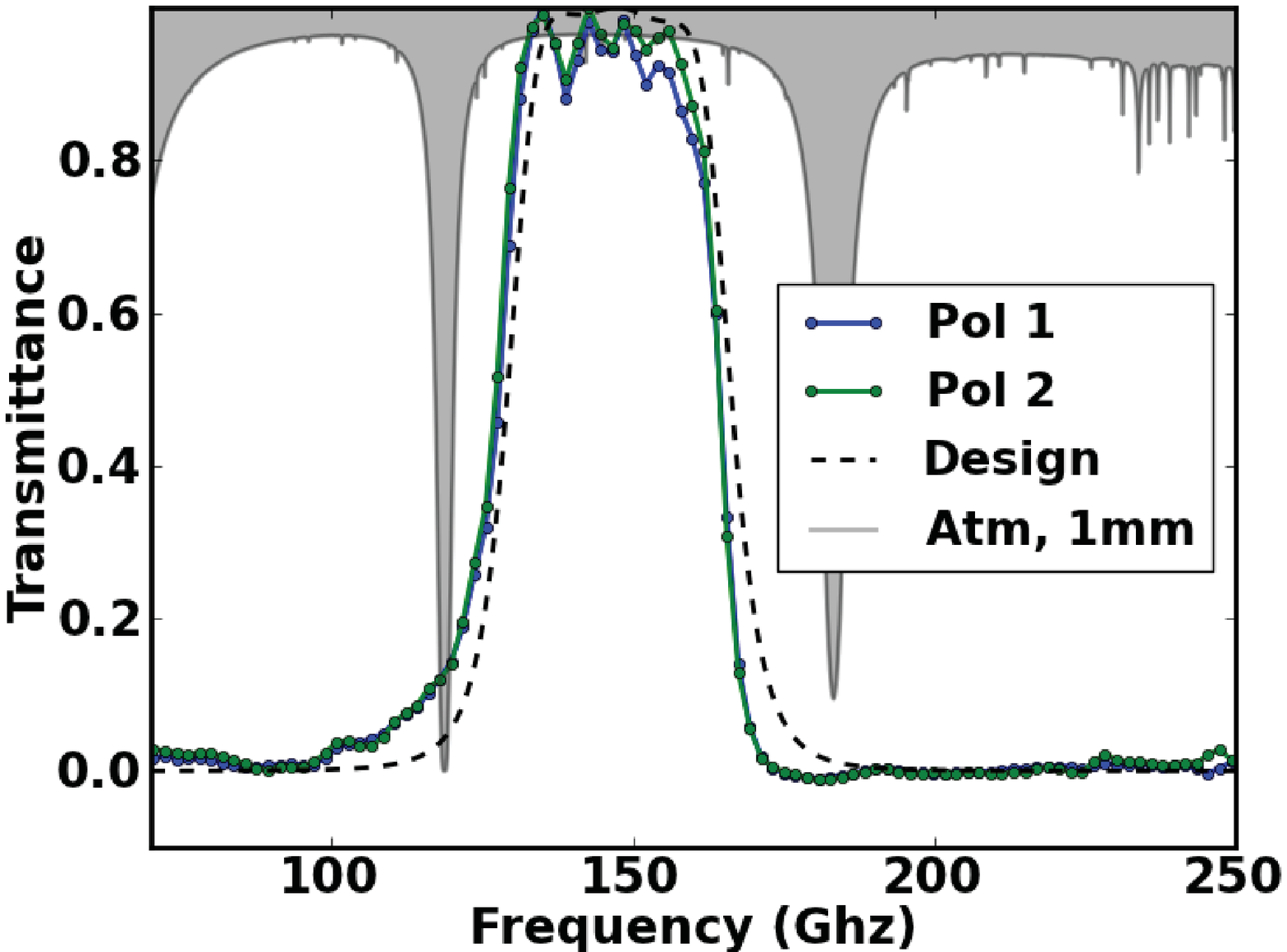} \\
\textbf{(a)} &\textbf{(b)}
\end{tabular}
	\end{center}
	\caption[Polariztion purity]{\label{fig:polpurity} \textbf{(a)} Boresight cross-polar response measurement of the individuel detectors in a \pb\ pixel. The figure shows data and a fit sinusoidal model with parameters for amplitude, phase, and cross-polar response. The fit cross-polar response is less than 1\% for each bolometer. \textbf{(b)} The spectral response of each bolometer for a single \pb\ pixel, plotted with the design spectral band and a model of the atmospheric transmittance in Chile for 1 mm of precipitable water vapor. The spectra are well-matched to each other and to the design band.}
	\end{figure}

A modulated thermal source was also used to measure the pixel's boresight cross-polar response. The thermal source was placed at the location of peak spatial response, and a lithographed wire grid polarizer on a thin dielectric sheet was placed between the thermal source and the detector. The response of both bolometers in the pixel was recorded as the polarizer was rotated. A model was fit to the data of the form $c_0 +\left ( c_1-c_0 \right ) \cos^2\left (\theta-c_3\right )$. The cross-polar response for each bolometer is defined as $c_0/c_1$, and was found to be less that 1\%. Figure \ref{fig:polpurity}\textbf{(a)} shows one such measurement.

\subsection{Spectral response}

The spectral response of the \pb\ detectors was measured in the lab using a \ac{fts}. Repeatable spectral measurements required that the spatial beam of the detector be filled by the output of the spectrometer. This was achieved using a warm UHMWPE lens to match the spectrometer output to the optics of the observation cryostat. Figure \ref{fig:polpurity}\textbf{(b)} shows the spectral band measurement for one of the pixels in the \pb\ focal plane array, and \tabref \ref{tab:spectra} enumerates the average spectral band parameters for each of the wafers in the array. Six of the seven average spectral bands are within the goal 5\% of band central location and integrated bandwidth.

\begin{table}
\vspace{6pt}
\begin{center}
\begin{tabular}{c|cc}
\textbf{Wafer} & \textbf{\shortstack{Center\\frequency\\(GHz)}} & \textbf{\shortstack{Integrated\\bandwidth\\(GHz)}} \\
\hline
Design & 148 & 38 \\
8.2.0 & 139 & 31 \\
9.4 & 148 & 36 \\
10.1 & 141 & 37 \\
10.1 & 145 & 38 \\
10.1 & 150 & 35 \\
10.4 & 145 & 36 \\
10.5 & 146 & 36
\end{tabular}
\end{center}
\caption[Spectral properties]{\label{tab:spectra}Per-wafer average measured center frequencies and integrated bandwidths. Design values are 148 GHz and 38 GHz, respectively.}
\end{table}



\subsection{Bolometer yield}

An important factor in total array sensitivity is the number of bolometers that are functional in the deployed array. Of the 1274 antenna-coupled bolometers in the \pb\ focal plane, 1015 show nominal optical response to a planet, an optical yield of 80\%. The factors that affect this number are shown in \tabref \ref{tab:yield}.

\subsection{Bolometer optical time constants}

\begin{figure}[htpb]
   \begin{center}
   \includegraphics[width=3.5in]{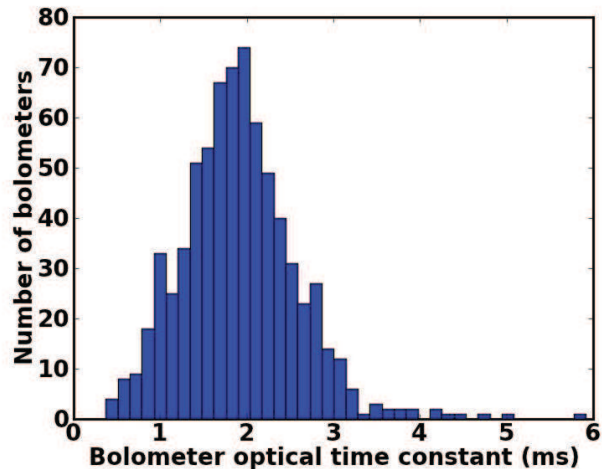}
	\end{center}
	\caption[Electrothermal properties]{\label{fig:electrothermal} Detector optical time constants as measured during operation. Peak value of 1.9 ms corresponds to an optical bandwidth of 84 Hz, and is fast enough to measure the changing \ac{cmb} signal as the \pb\ telescope scans across it.}
	\end{figure}

The detector optical time constants were measured under observing conditions using a modulated thermal source that is instrumented through a 6 mm hole in the secondary mirror. A single pole time constant model is a good fit to the detectors' response functions, and the uncertainty in the fit for most detectors is less than 0.2 ms. A histogram of the measured optical time constants, with a peak value around 1.9 ms, is shown in \figref \ref{fig:electrothermal}. This is fast enough to measure the changing \ac{cmb} signal as the \pb\ telescope scans across it.

\begin{table}
\begin{center}
\begin{tabular}{lr}
\textbf{Functional check} & \textbf{Number} \\
\hline
Total fabricated (1274 antenna-coupled, 70 ``dark'') & 1344 \\
Passed warm inspection & 1245 \\
Electrically responsive cold & 1154 \\
Biasable and shows nominal beammaps & 1015
\end{tabular}
\vspace{6pt}
\caption[Bolometer yield]{\label{tab:yield}Bolometer yield in the deployed \pb\ focal plane. Of the 1274 antenna-coupled bolometers (70 are not coupled to antennas and are used as checks for spurious signals), $1015/1274=80$\% show nominal beammaps.}
\end{center}
\end{table}

\section{THE FUTURE OF POLARBEAR AND LENSLET-COUPLED DETECTORS}
\label{sec:future}

\Pb\ is currently observing from Cerro Toco in Northern Chile, which marks the \pb\ detector array as the first lenslet-coupled detectors to observe the \ac{cmb}. This is an exciting advance, both because lenslet-coupling proves a path for a multi-band pixel architecture that will be an enabling technology for the next generation of \ac{cmb} experiments\cite{Suzuki_SPIE2012}, and because the current \pb\ instrument will improve our understanding of cosmology with the data it collects in the coming years.

\acknowledgments     
 
The \pb\ project is funded by the National Science Foundation under grant AST-0618398. Antenna-coupled bolometer development at Berkeley is also funded by NASA under grant NNG06GJ08G. The KEK authors were supported by MEXT KAKENHI Grant Number 21111002. The McGill authors acknowledge funding from the Natural Sciences and Engineering Research Council and Canadian Institute for Advanced Research. MD acknowledges support from an Alfred P. Sloan Research Fellowship and Canada Research Chairs program. All silicon wafer-based technology is fabricated at the UC Berkeley Nanolab.


\bibliography{master,myAuthorship}
\bibliographystyle{spiebib}   

\end{document}